\documentclass[runningheads]{llncs}
\usepackage{graphicx}
\usepackage{relsize}
\graphicspath{./figures/}
\usepackage{subcaption}
\usepackage{amsmath}
\usepackage{amsfonts}
\usepackage{booktabs}
\usepackage[colorinlistoftodos,prependcaption,textsize=tiny]{todonotes}

\def \Rm{{\mathbb{R}}}

\def \Abf{{\mathbf A}}

\def \Dbf{{\mathbf D}}

\def \Hbf{{\mathbf H}}

\def \Wbf{{\mathbf W}}

\def \Zbf{{\mathbf Z}}
\def \Hbf{{\mathbf H}}
\def \0bf{{\mathbf 0}}

\begin{document}
\title{A joint 3D UNet-Graph Neural Network-based method for Airway Segmentation from chest CTs}
\titlerunning{Abbreviated paper title}
%
\author{Antonio Garcia-Uceda Juarez\inst{1,2}, Raghavendra Selvan\inst{3}, Zaigham Saghir\inst{4} \and Marleen de Bruijne\inst{1,3}
}
\titlerunning{A 3D UNet-Graph Neural Network method for Airway Segmentation}
\authorrunning{Garcia-Uceda Juarez et al.}
%
\institute{Biomedical Imaging Group Rotterdam, Department of Radiology and Nuclear Medicine, Erasmus MC, 3015 CE, Rotterdam, The Netherlands \\ \email{a.garciauceda@erasmusmc.nl} \and Department of Pediatric Pulmonology, Erasmus MC-Sophia Children Hospital, 3015 CE, Rotterdam, The Netherlands \and Department of Computer Science, University of Copenhagen, DK-2100, Copenhagen, Denmark \and Department of Medicine, Section of Pulmonary Medicine, Herlev-Gentofte Hospital, Copenhagen University Hospital, Kildegårdsvej 28, 2900 Hellerup, Denmark
}
\vspace{300mm}
\maketitle              
\begin{abstract}
We present an end-to-end deep learning segmentation method by combining a 3D UNet architecture with a graph neural network (GNN) model. In this approach, the convolutional layers at the deepest level of the UNet are replaced by a GNN-based module with a series of graph convolutions. The dense feature maps at this level are transformed into a graph input to the GNN module. The incorporation of graph convolutions in the UNet provides nodes in the graph with information that is based on node connectivity, in addition to the local features learnt through the downsampled paths. This information can help improve segmentation decisions. By stacking several graph convolution layers, the nodes can access higher order neighbourhood information without substantial increase in computational expense. We propose two types of node connectivity in the graph adjacency: i) one predefined and based on a regular node neighbourhood, and ii) one dynamically computed during training and using the nearest neighbour nodes in the feature space. We have applied this method to the task of segmenting the airway tree from chest CT scans. Experiments have been performed on 32 CTs from the Danish Lung Cancer Screening Trial dataset. We evaluate the performance of the UNet-GNN models with two types of graph adjacency and compare it with the baseline UNet.
\end{abstract}
\section{Introduction}
\label{sec:Introduction}
\vspace{-0.2cm}
Since recently, fully convolutional neural networks (CNNs) are the state-of-the-art for many segmentation tasks~\cite{LongCVPR2015}, and in particular the UNet architecture~\cite{RonnenMICCAI2015} for biomedical image segmentation. The UNet consists of an encoding path, in which high-order features are extracted at several downsampled resolutions, followed by a decoding path, in which these features are leveraged to the full resolution to perform voxel-wise segmentation decisions. An extension of CNNs to graph structured data are Graph neural networks (GNNs)~\cite{Scarselli2009GNN,Kipf2017Semi}, which have seen early application for segmenting structures that resemble graphs~\cite{SelvanMIDL2018,ShinArxiv2018}. Initial work of combining CNNs and GNNs was by proposed Shin et al.~\cite{ShinArxiv2018} for 2D vessel segmentation. This was a sequential pipeline in which the CNN was trained for feature extraction prior to applying the GNNs to learn global connectivity.

The segmentation of tree-like structures such as the airways in chest CTs is a complex task, with branches of varying sizes and different orientations while taking into account bifurcations. A comparison of airway extraction methods (prior to CNNs) was performed in the EXACT challenge~\cite{LoExact2010}. The results showed that all methods missed a significant amount of the small, peripheral branches. The UNet has since been applied for airway segmentation in~\cite{MengMICCAI2017,GarciaMICCAI2018}. The 3D UNet-based method in~\cite{GarciaMICCAI2018} is fully automatic and can segment the airways in a full lung in a single pass through the network. However, this method had problems to capture various small terminal branches. Also, the GNNs have been applied to airway extraction as a graph refinement approach in~\cite{SelvanMIDL2018}. However, this method was not end-to-end optimised and relied on handcrafted features as input to the graph.

In this paper, we present an end-to-end segmentation approach by combining a 3D UNet with GNNs and evaluate this on extracting the airway tree from chest CTs. This method replaces the two convolutional layers in the deepest level of the baseline UNet by a GNN-based module, comprising graph convolutions. This end-to-end approach simultaneously learns local image features and global information based on graph neighbourhood connectivity.
\vspace{-0.1cm}
\section{Methods}
\label{sec:Methods}
\begin{figure}[t]
\centering
\begin{subfigure}{0.7\textwidth}
\includegraphics[width=1.0\linewidth]{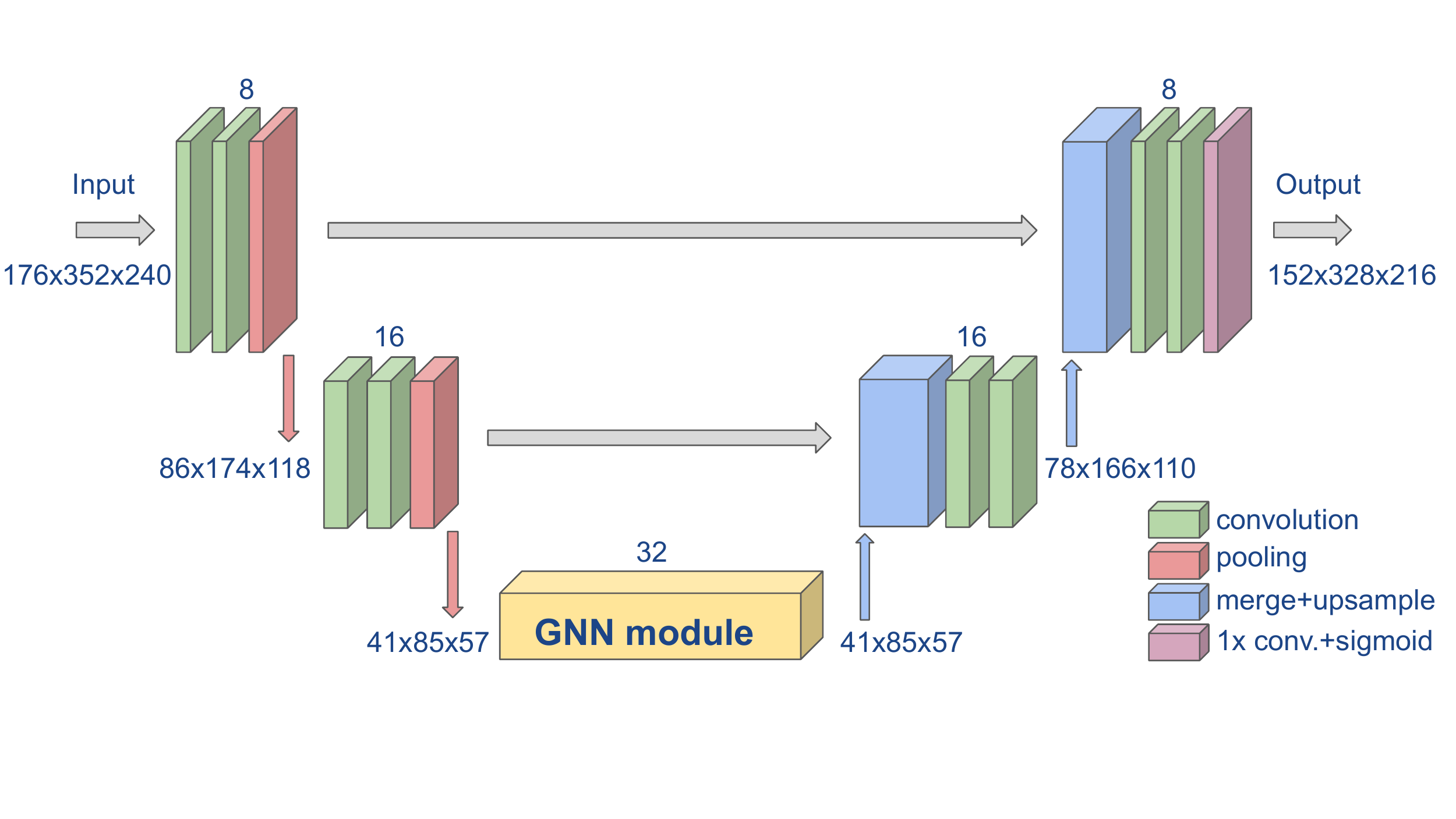}
\end{subfigure}
\begin{subfigure}{0.25\textwidth}
\includegraphics[width=1.0\linewidth]{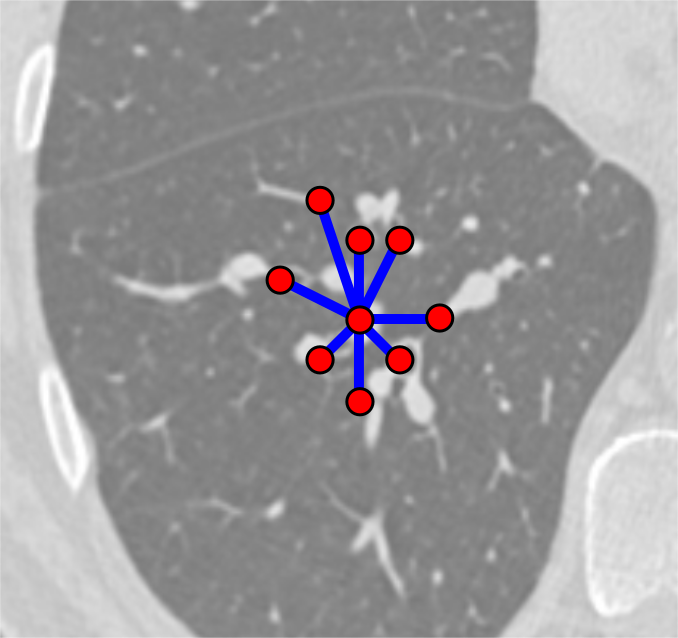}
\end{subfigure}
\vspace{-0.7cm}
\caption{Left: Schematics of a UNet-GNN network of 3 levels. Right: Illustration of irregular node connectivity for a given voxel in the initial graph.} \label{fig1}
\vspace{-0.3cm}
\end{figure}

\vspace{-0.2cm}
The proposed joint UNet-GNN architecture is described in the following subsections. This approach integrates a GNN module at the deepest level of a baseline 3D UNet, and is schematically shown in Fig.~\ref{fig1} (left). The GNN module uses a graph structure obtained from the dense feature maps resulting from the contracting path of the Unet. Each graph node can be viewed as a "supervoxel" from the downsampled regions with the corresponding vector of features. The connectivity of nodes in such a graph is described by the adjacency matrix and determines the neighbourhood of each node when performing graph convolutions, as shown in Fig.~\ref{fig1} (right). The GNN module learns combinations of the input feature maps based on the graph topology, and outputs another graph with same "supervoxel" nodes as the input graph and the corresponding vector of learnt features for each node. This output are feature maps that are fed to the upsampling path of the UNet.

We have tested two types of node connectivity that are: i) Predefined and based on a regular neighbourhood, with each node connected to its 26-direct neighbours, and ii) Dynamically computed during training based on choosing nearest neighbours in the node feature space. Stacking several GNN layers allows nodes to access longer range information beyond the initial neighbourhood, which can improve the segmentation decisions as more complex features that include relevant information from nodes far away in the volume can be used. This is in contrast with CNNs which rely on local feature extraction, and access long range information via successive convolutions and pooling, with a detriment of strong reduction in resolution. This long range access by the GNN module is useful when segmenting tree structures like airways, as information from branches that share some features (shape, orientation, ...) but are further away in the volume can be used when detecting a given branch. The computation of the dynamic graph adjacency is further explained in Section~\ref{sec:IrregNeighs}. 

\vspace{-0.2cm}
\subsection{Graph Neural Network (GNN) module}
\label{sec:GNNmodule}
\vspace{-0.2cm}
The main component of the GNN module is a series of graph convolutional layers~\cite{Scarselli2009GNN,Kipf2017Semi}. This operation can be seen as a generalisation of the cartesian convolution to a graph setting. One of the formulations of graph convolution operation with separate processing of self connections proposed in Kipf et al.~\cite{Kipf2017Semi} is given by the equation,
\begin{equation}
\Hbf^{(l+1)} = \mathlarger\sigma\Big( \Hbf^{(l)}\Wbf_0^{(l)}+\Dbf^{-1}\Abf \Hbf^{(l)}\Wbf_1^{(l)}\Big)
\label{eq:gnn}
\end{equation}
where  $\mathlarger\sigma(\cdot)$ is the rectified linear unit activation function, $\Hbf^{(l)} \in \Rm^{N \times E} $ is the node feature matrix comprising $N$ nodes and $E$ features input to the $l^{\text{th}}$ GNN layer. The learnable GNN filter weights are $\Wbf^{(l)}_0, \Wbf^{(l)}_1 \in \Rm^{E\times E}$.  $\Abf \in \{0,1\}^{N\times N}$ is the binary adjacency matrix, $\Dbf$ is the degree matrix derived from $\Abf$ with diagonal entries $D_{ii} = \sum_{j=1}^{N} A_{ij}, \forall i=1\dots N $. The adjacency matrix is largely sparse, with non-zero entries per node corresponding to the size of each node neighbourhood (26 for the regular neighbourhood case above). By processing the adjacency matrix as a sparse tensor, operations in Eq.~\eqref{eq:gnn} are done efficiently. 

The GNN module in the proposed method has $L=4$ layers performing the operations in Eq.~\eqref{eq:gnn} successively. By stacking several graph convolutions together, each node in the output graph updates its features with information from higher order neighbourhood, which can improve the segmentation decisions. The initial graph features $\Hbf^{(0)}$ are obtained as,
\begin{equation}
    \Hbf^{(0)} = f(\Zbf)
    \label{eq:nodeMLP}
\end{equation}
where $\Zbf \in \Rm^{N\times F} $ is the F-dimensional node feature matrix derived from the UNet and $f(\cdot)$ is a two layered multi-layer perceptron with $F$ input units and $E$ output units and rectified linear units, followed by a normalization layer. The transformation in Eq.~\eqref{eq:nodeMLP} allows more complex representations of the input node features useful for the GNN module.
\vspace{-0.2cm}
\subsection{Irregular Neighbourhood}
\label{sec:IrregNeighs}
\vspace{-0.2cm}
A GNN module with regular adjacency has limited node connectivity comprising only the direct neighbours. This constraint is imposed primarily to keep the adjacency matrix sparse to reduce the large memory footprint. In order to allow nodes in the graph to access information well beyond their directly connected neighbours, we propose an extension to the GNN module in Section~\ref{sec:GNNmodule} by using a graph adjacency with node connectivity as the $k-$nearest neighbours in the feature space. Node neighbourhood is decided from the pairwise euclidean distance between nodes from $\Zbf$ in the $F-$ dimensional feature space. In this work, we set the number of neighbours to $k=26$. The graph adjacency is dynamically computed during training for every input image in every epoch. This approach, we argue, enables the method to access irregular but meaningful neighbourhoods and utilises the inherent capabilities of using GNN-based learning over irregular neighbourhoods in an improved fashion.
\vspace{-0.2cm}
\section{Experiments}
\label{sec:Experiments}
\vspace{-0.2cm}
\subsection{Dataset}
\label{sec:Dataset}
\vspace{-0.2cm}
We used 32 low-dose CT chest scans from the Danish Lung Cancer Screening Trial~\cite{PedersenDLCST2009}. All scans have voxel-resolution of roughly 0.78x0.78x1 mm$^3$. The reference segmentations are airway lumen obtained from the method~\cite{PetersenMIA2014} applied on the union of two previous methods~\cite{LoMIA2010,LoMICCAI2009}, and corrected by an expert observer.
\vspace{-0.4cm}
\subsection{Network Implementation}
\label{sec:NetImpl}
\vspace{-0.2cm}
The baseline UNet upon which the UNet-GNN approach is built derives from the model in~\cite{GarciaMICCAI2018} with 3 levels of resolution. Each level in the downsampling / upsampling path is composed of two 3x3x3 convolutional layers with rectified linear (ReLU) activation, and followed by a 2x2x2 pooling / upsample layer, respectively. No padding is used in the convolutions in order to reduce the model memory footprint. There are 8 feature maps in the first level, and this is doubled / halved after every pooling / upsampling, respectively. The GNN module in the deepest level has twice as many output features per voxel as input units, i.e. $E=2F$ in Eqs.~\eqref{eq:gnn} and~\eqref{eq:nodeMLP}. The final layer consists of a 1x1x1 convolutional layer followed by a sigmoid activation function.

A series of operations to extract the input images from the volume CTs as in~\cite{GarciaMICCAI2018} are applied. These consist of: i) crop the CTs to a bounding-box around pre-segmented lung fields, ii) extract smaller input images in a sliding-window fashion in axial dimension, and iii) apply image rigid transformation for data augmentation. The model is designed for the largest input images that can fit in GPU memory. This corresponds to a size 176x352x240 in a batch containing only one image, for a GPU NVIDIA GeForce GTX 1080 Ti with 11 GB memory used in these experiments. The models were implemented in Pytorch framework.
\vspace{-0.7cm}
\subsection{Training the models}
\label{sec:TrainModels}
\vspace{-0.2cm}
The loss function used for training the network is the dice loss,
\begin{equation}
\mathcal{L} = \frac{2 \sum_{x\in N_L}{p(x) g(x)}}{\sum_{x\in N_L}{p(x)} + \sum_{x\in N_L}{g(x)} + \epsilon} 
\end{equation}
where $p(x), g(x)$ are the predicted voxel-wise airway probability maps and airway ground truth, respectively. The ground truth is masked to the region of interest (ROI): the lungs, indicated by the sub-index $L$, so that only voxels within this region contribute to the loss. This mask removes the trachea and main bronchi from the ground truth, so that training is focused on the smaller branches. The lung segmentation needed for this masking operation is easily computed by a region-growing method in~\cite{LoMIA2010}. $\epsilon$ is a tolerance needed when there are no ground truth voxels in the image patch.

To train the networks we use 16 CTs, 4 CTs for model validation / hyperparameter tuning, and the remaining 12 CTs for testing. The Adam optimizer is used with a learning rate that is chosen for each model as large as possible while ensuring convergence of the losses. This was $10^{-4}$ for UNet-GNN models and $5\times10^{-5}$ for UNet models. All models are trained until convergence, and we retrieve the model with overall minimum validation loss for testing. As convergence criterion, we monitor the moving average of the validation loss over 50 epochs, and training is stopped when its value i) increases by more than 5\%, or ii) does not decrease more than 0.1\%, over 20 epochs (patience). Training time was approximately 1-2 days, depending on the models, while test time inference takes a few seconds per scan.
\vspace{-0.3cm}
\subsection{Experiments set-up}
\label{sec:ExpSetup}
\vspace{-0.2cm}
We evaluate four different models: two regular UNets with 3 resolution levels (UNetLev3) and 5 levels (UNetLev5), and two UNet-GNN models each with 3 levels. The UNet-GNN models differ in the type of graph neighbourhood: i) regular graph adjacency with 26 direct neighbours (UGnnReg), and ii) dynamic computation of the adjacency matrix (UGnnDyn), with 26 connections per node as described in Section~\ref{sec:IrregNeighs}. With our experiments we evaluate: i) the benefit of the GNN module at deepest level of the UNet when compared to the two convolutional layers of a regular UNet, and ii) the difference in performance of UNet-GNN models when compared to a more complex model like the UNetLev5.

The models are compared based on i) Dice overlap coefficient, ii) airway completeness, iii) volume leakage, and two centreline distance error measures, iv) false negative $d_{FN}$ and v) false positive $d_{FP}$ distances, as in~\cite{SelvanMIDL2018}. Completeness is the ratio of ground truth centreline length inside the predictions with respect to ground truth centreline full length, and volume leakage is the ratio of number of false positive voxels with respect to the number of voxels in the ground truth. We refer the reader to~\cite{SelvanMIDL2018} for the definition of $d_{FN}$ and $d_{FP}$. Trachea and main bronchi are removed in these measures from both predictions and ground truth.

We compute the ROC curves for all the models using the mean airway completeness and volume leakage measured on the test set, and varying the threshold in the output probability maps used to obtain the final binary segmentations. Further, we compute the operating point with a fixed level of volume leakage (13\%) by estimating the correct threshold (with a minimum error $10^{-4}$). This threshold is, for the different models: UNetLev3 (0.1), UNetLev5 (0.04), UGnnReg (0.66), UGnnDyn (0.33). We use the resulting airway segmentations to evaluate the performance measures on the test set. To compare the models, we use the two-sided T-test on two related distributions.

The calculation of the dynamic graph adjacency as described in Section~\ref{sec:IrregNeighs} is computationally expensive, both in memory and run time. In order to fit the model in GPU and reduce the computational cost, the searching space of candidate nodes to compute the node connectivity is constrained to a cube area of at most 5 voxels away from the given node. This limits the range for direct node connections, however long range information is still accessed via the four stacked graph convolutions as described in Section~\ref{sec:GNNmodule}. 
\vspace{-0.2cm}
\section{Results and Discussion}
\label{sec:Results}
\vspace{-0.8cm}

\begin{figure}[h]
\centering
\includegraphics[width=0.49\linewidth]{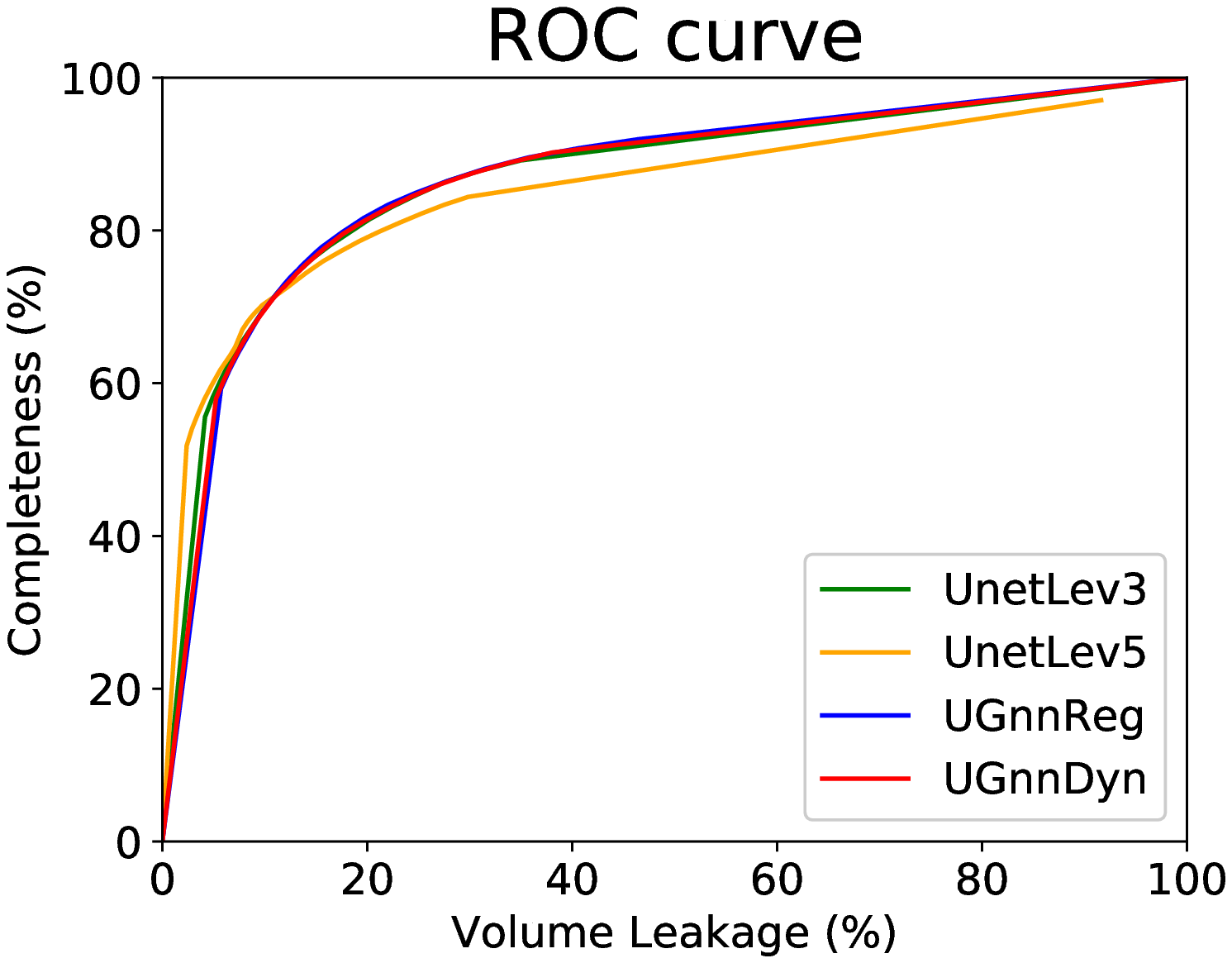}
\includegraphics[width=0.49\linewidth]{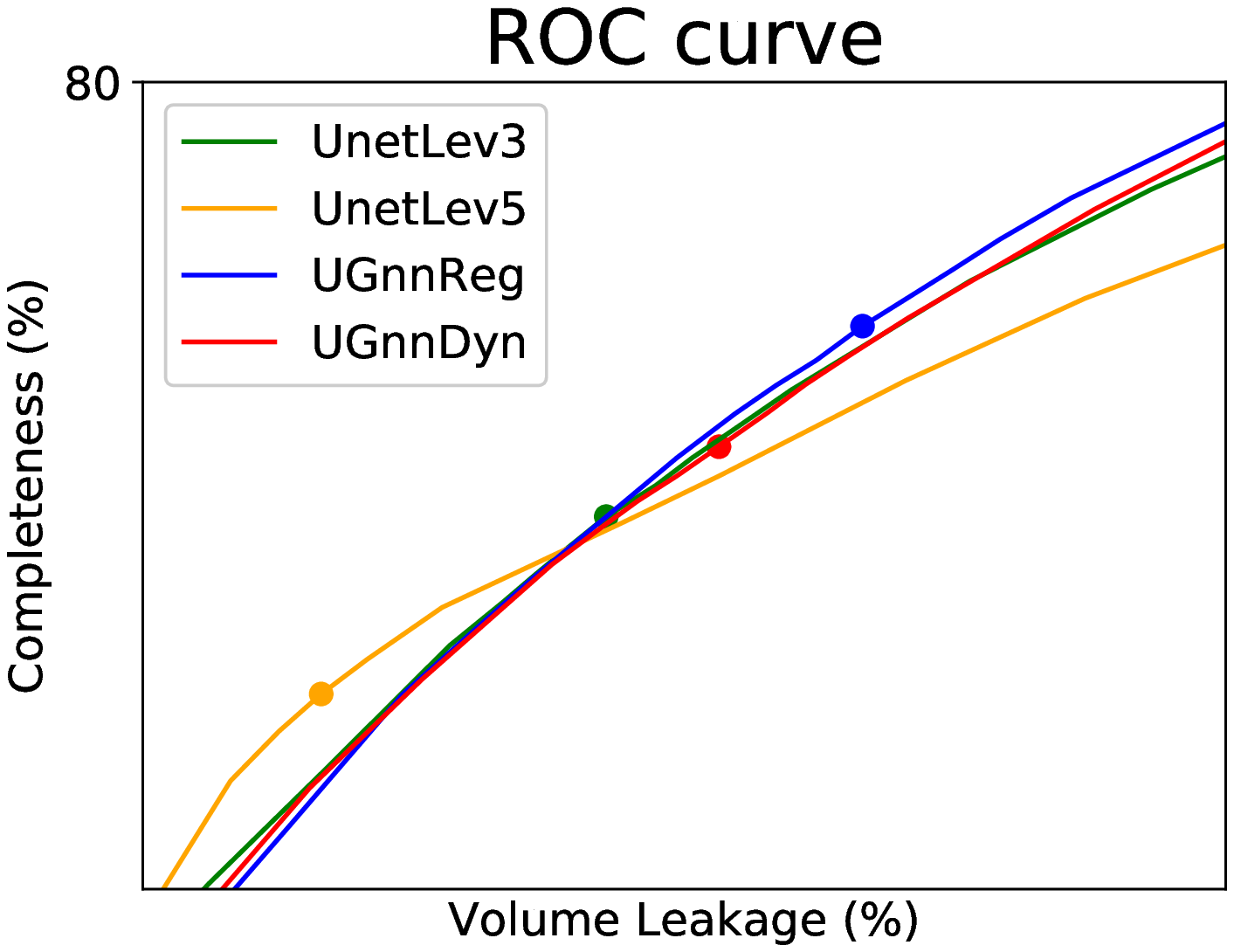}
\caption{ROC curves for all the models, varying the threshold in the probability maps. Right: detailed view, including the operating points with threshold 0.5.}\label{fig2}
\vspace{-0.5cm}
\end{figure}

\begin{figure}[h]
\begin{subfigure}{1.0\textwidth}
\centering
\includegraphics[width=0.32\linewidth]{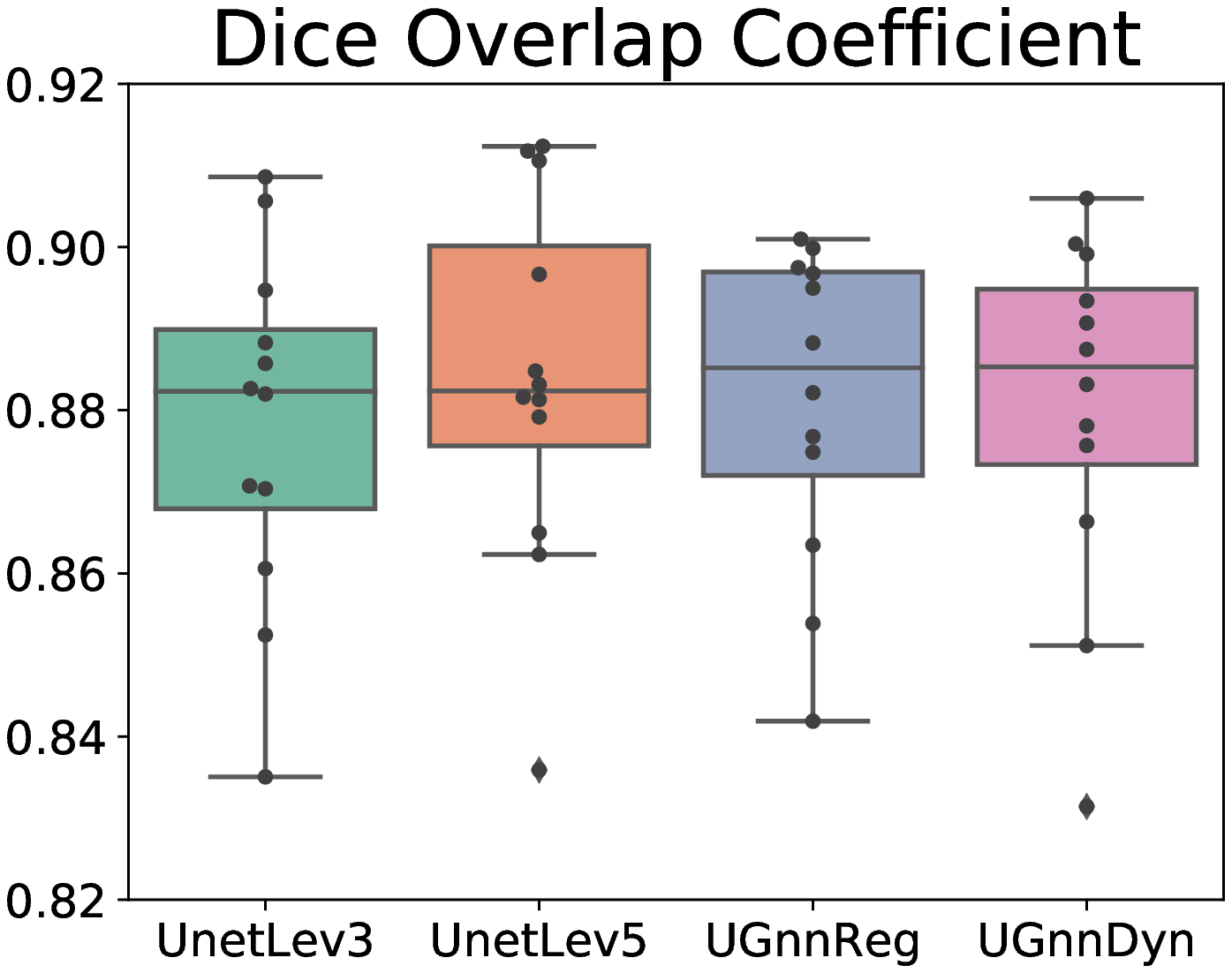}
\includegraphics[width=0.32\linewidth]{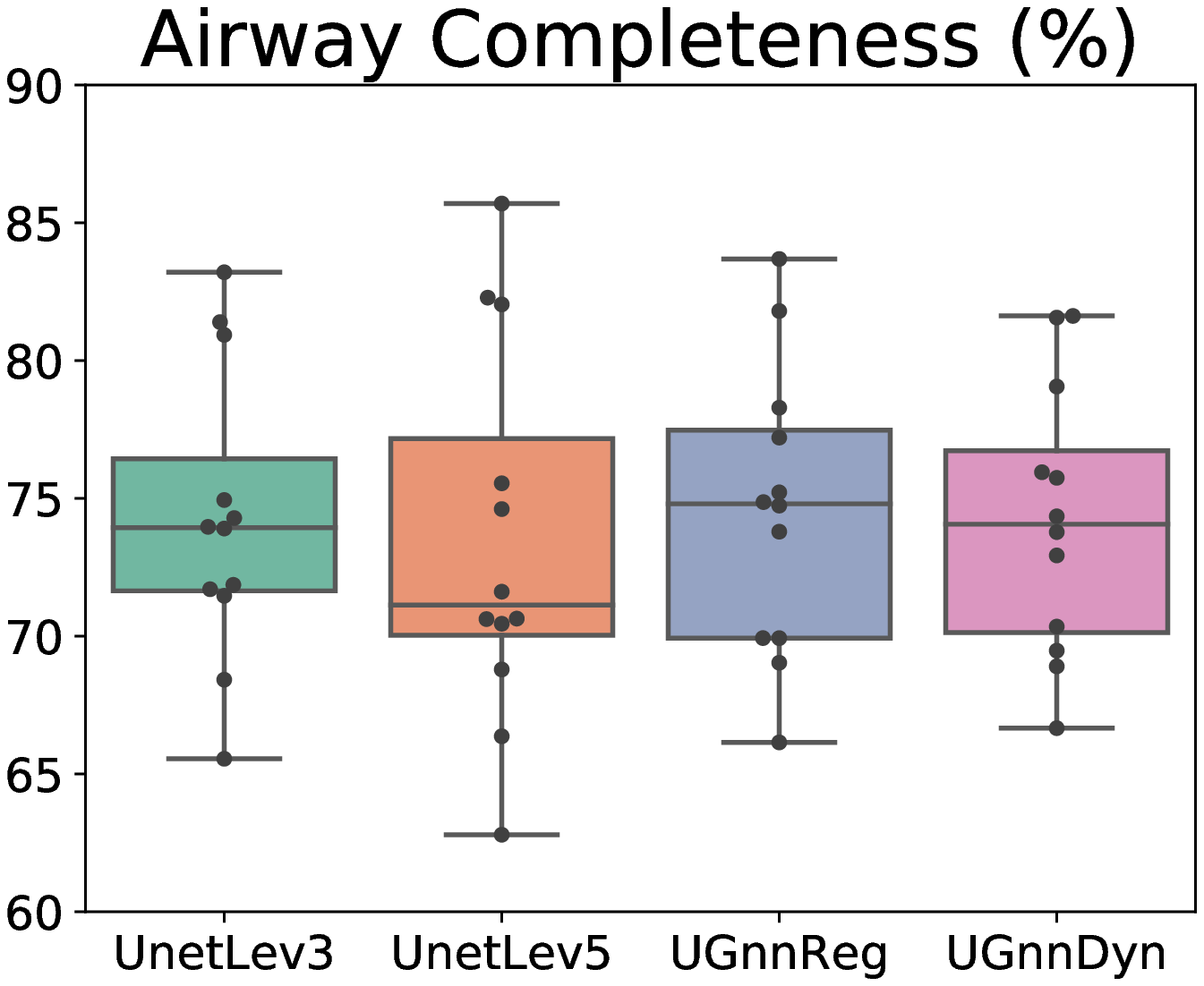}
\includegraphics[width=0.32\textwidth]{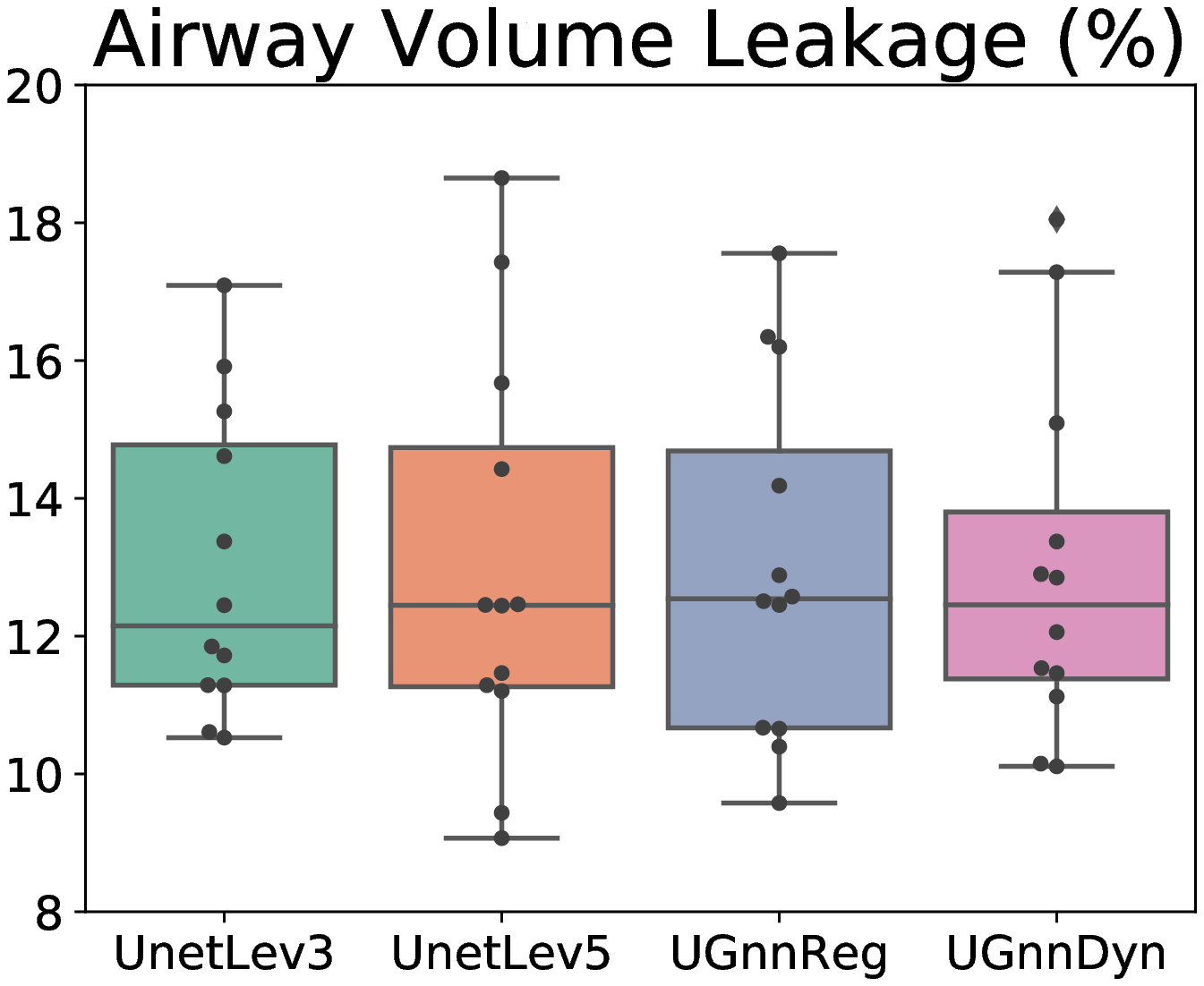}
\end{subfigure}
\begin{subfigure}{1.0\textwidth}
\centering
\includegraphics[width=0.4\linewidth]{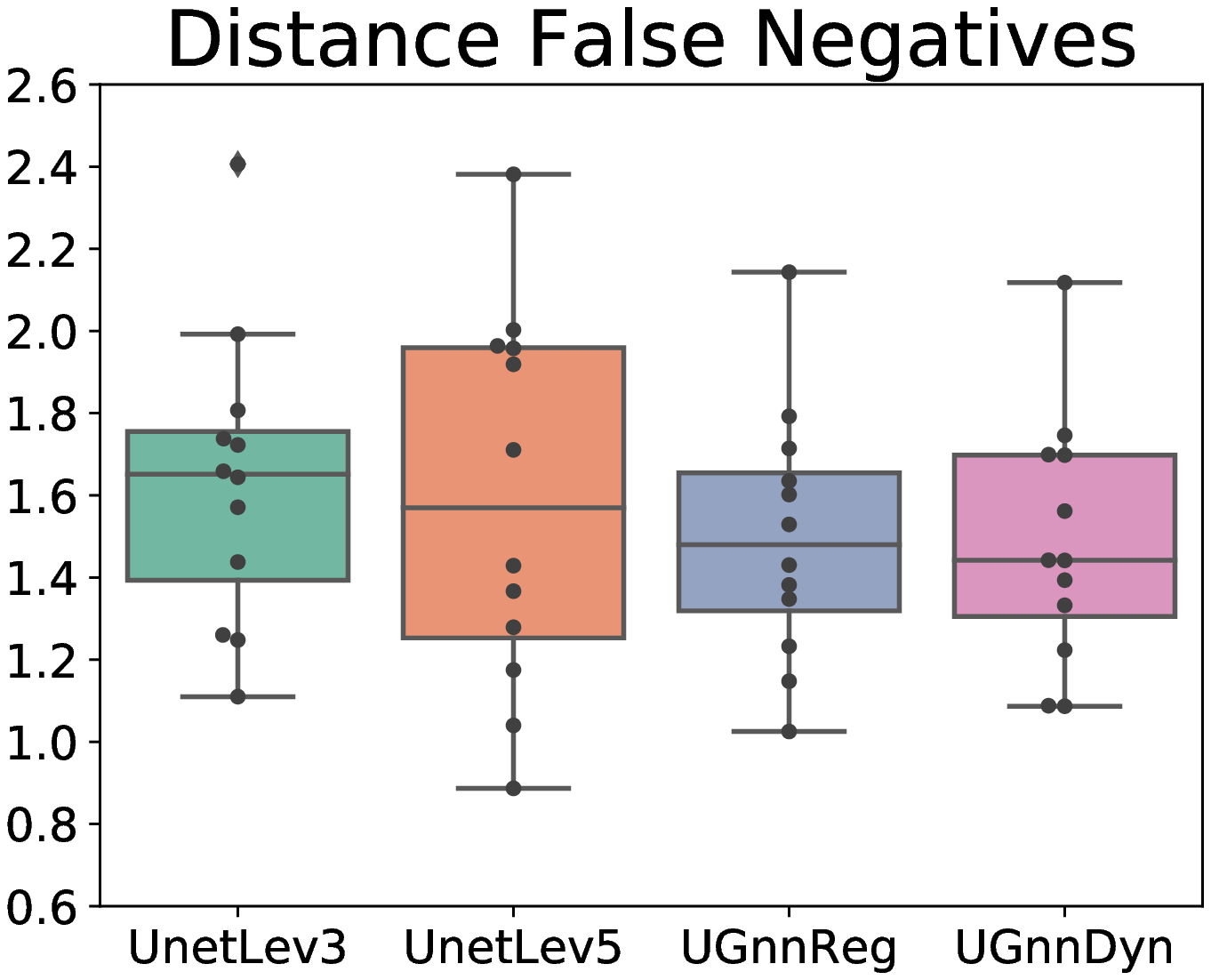}
\includegraphics[width=0.4\linewidth]{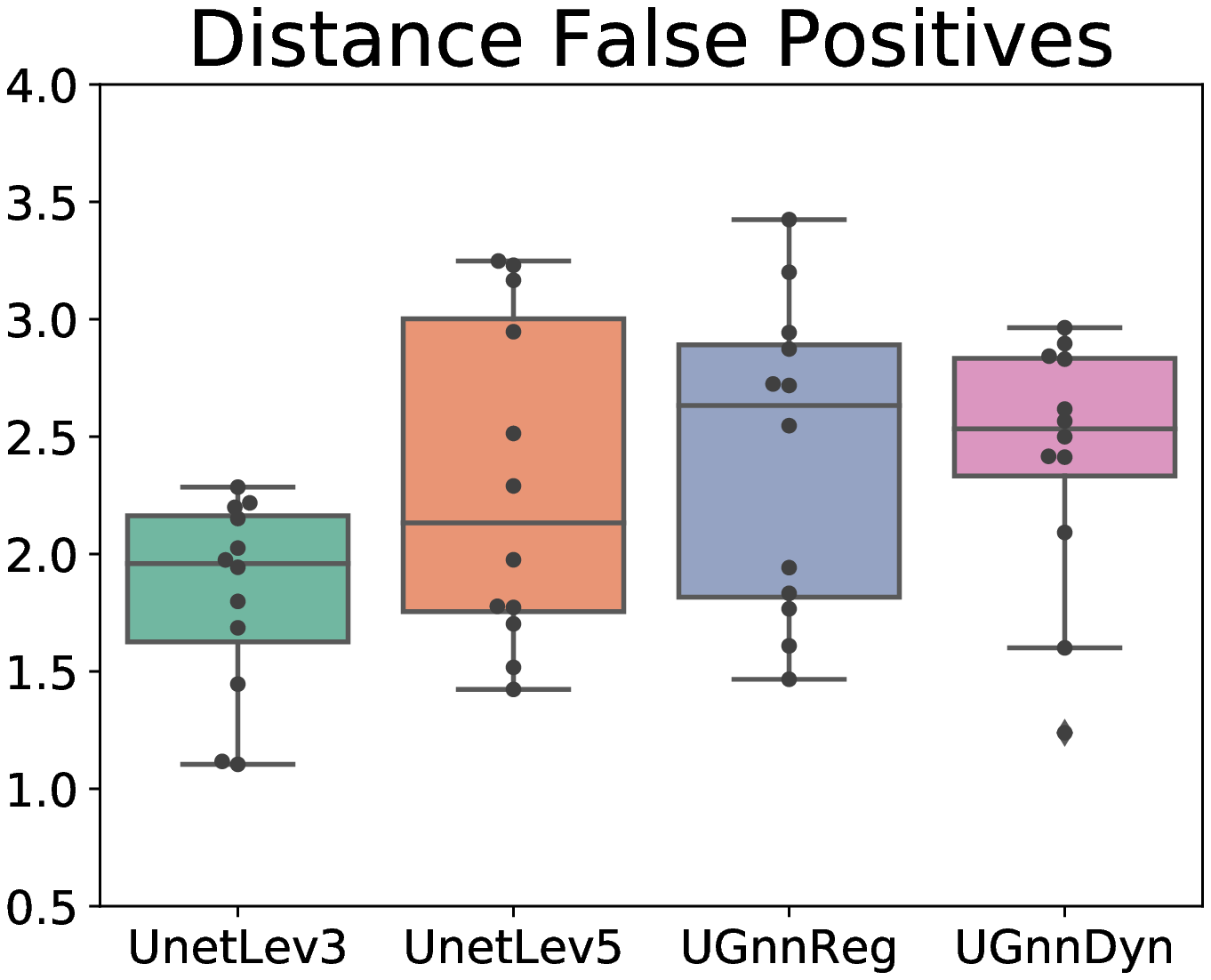}
\end{subfigure}
\caption{Test set performance measures for the different models.}\label{fig3}
\vspace{-0.5cm}
\end{figure}

The ROC curves for the different models are shown in Fig~\ref{fig2}. The proposed UNet-GNN models and the baseline UNetLev3 show very similar results. In the detailed view, the UGnnReg shows a small improvement, i.e. higher airway completeness for a given volume leakage, over the baseline UNetLev3. Further, the UnetLev5 shows lower completeness than the other models for volume leakage higher than 11\%, while the contrary occurs for lower leakage levels. 

Performance of the different models on the test set, for an operating point in the ROC curve with 13\% volume leakage, is shown in Fig~\ref{fig3}. In dice overlap coefficient, airway completeness and volume leakage, there is no significant difference between the models ($p>0.1$). Nevertheless, in centreline distance measures, the proposed UNet-GNN models show significantly lower $d_{FN}$ and higher $d_{FP}$ with respect to the baseline UNetLev3: in $d_{FN}$, comparing UGnnDyn ($p=0.001$) and UGnnReg ($p<0.01$) with UNetLev3, and in $d_{FP}$, comparing both UNet-GNN models with UNetLev3 ($p<0.001$). Further, the UNetLev5 shows no significant difference in $d_{FP}$ and $d_{FN}$ with the other models ($p>0.1$).

The small and significant improvement in distance false negative error $d_{FN}$, with an on par dice overlap and airway completeness, indicates that the proposed UNet-GNN models can segment slightly more complete airway trees, when compared to the two UNet models, with more and/or longer peripheral branches. This is because the centreline distance measure $d_{FN}$ is not dependent on the scale of airways and provides a uniform measure of accuracy in detecting branches. In contrast, the completeness measure is a binary evaluation of branches with respect to the reference and small improvements in centreline detection may not be reflected in the overall score. On the contrary, the more complex model UNetLev5 does not show any difference in any measure over the baseline UNetLev3. Further, the UNet-GNN models have fewer trainable parameters ($\approx 50k$) than the UNetLev3 ($\approx 90k$) and much less than the UNetLev5 ($\approx 1.4M$).

One aspect that might limit the performance of the proposed UNet-GNN models is that the GNN module is operating only on the deepest level of the UNet, where the image patches have undergone three downsampling operations. A more powerful UNet-GNN model can be formulated by replacing all skip connections in the baseline UNet with GNN modules, each operating at that image resolution level. However, this model has not been experimented with due to the large memory footprint required, which exceeded the available GPU resources at our disposal.
\vspace{-0.2cm}
\section{Conclusions}
\label{sec:Conclusions}
\vspace{-0.2cm}
We presented a joint UNet-GNN based segmentation method with an application to extract airways from chest CTs. By introducing a GNN module with graph convolutions at the deepest level of the UNet the proposed method is able to learn and combine information from a larger region of the image. The proposed UNet-GNN models show a small and significant improvement in the false negative centreline measure over the baseline UNet, with on par dice overlap and airway completeness, for a fixed volume leakage. This indicates that the proposed UNet-GNN models can segment slightly more complete airway trees. Further, this is achieved with fewer trainable parameters.
%
%
%

\end{document}